\documentclass[twocolumn,showpacs,preprintnumbers,amsmath,amssymb]{revtex4}
\usepackage{graphicx}
\usepackage{dcolumn}
\usepackage{bm}
\begin{document}
\preprint{APS/123-QED}
\title{The restrictions of classical mechanics in the description
of dynamics of nonequilibrium systems and the way to get rid of
them}
\author{V.M. Somsikov}
 \altaffiliation[] {}
 \email{nes@kaznet.kz}
\affiliation{%
Laboratory of Physics of the geoheliocosmic relation, Institute of
Ionosphere, Almaty, Kazakstan.
}%

\date{\today}
\begin{abstract}
The reasons which restrict opportunities of classical mechanics at
the description of nonequilibrium systems are discussed. The way
of overcoming of the key restrictions is offered. This way is
based on an opportunity of representation of nonequilibrium system
as a set of equilibrium subsystems. The equation of motion and the
general Lagrange, Hamilton and Liouville equations for subsystems
have been obtained. The way of a substantiation of thermodynamics
is offered.
\end{abstract}

\maketitle

\section{\label{sec:level1}Introduction\protect}

The development of the theoretical bases of physics of open
nonequilibrium systems is collided with contradictions between
classical mechanics and thermodynamics. These contradictions are
most brightly displayed in a problem of a substantiation of the
second law of thermodynamics or, in other words, explanations of
irreversibility. Since L. Boltzmann and to this day the attempts to
solve this problem are undertaken [1-4].

In the basis of the most widespread contemporary explanation of
irreversibility the mixing property inherent at the Hamiltonian
systems is used. The mixing leads to the irreversibility if one
postulates averaging of phase space on physically small volume.
But the explanation of the nature of such averaging within the
framework of classical mechanics is impossible [2, 3].

With the purpose of search of solutions of these problems the hard
disks systems were investigated [5, 6]. Its studying revealed that
if the nonequilibrium system of disks has been represented by
interacting equilibrium subsystems (IES) then equilibration is
caused by the work of collective forces of IES. These forces are
dependent on the disks velocities and their work transforms the
motion energy of IES into the internal energy [6].

For generalization of results of researches of hard disks it was
necessary to pass to the model of nonequilibrium systems of
potentially interacting of elements. Studying of disks systems has
shown that for this aim it is necessary to represent the model of
nonequilibrium system as a set of interacting IES. That model
possesses the big generality. Indeed, the base of statistical
physics was created by Gibbs using that model [7]. In particular
it has helped to introduce the distribution function into
classical mechanics basing on concept of probability of beginning
conditions for IES.

The method of Gibbs's ensembles is applicable for studying
equilibrium systems in the absence of an energy exchange between
IES. But the exchange of energy between IES in case of
nonequilibrium systems is responsible for an equilibration.
Nevertheless the splitting of nonequilibrium system of potentially
interacting elements into IES allows reducing the proof of
irreversibility to the proof of existence of non-potential part of
collective force between IES. It will be proved with the help of
the equation of interaction of systems (UVS) which was obtained
directly from the energy conservation law [8].

Basing on UVS and the model of nonequilibrium system as a set of
IES the new approach to the analysis of nonequilibrium systems has
been offered. The following assumptions and conditions are put
into the basis of this approach: 1). IES energy should be
submitted by the sum of internal energy and energy of IES motion
as a whole; 2). Each element of system should be fixed for
corresponding IES without dependence from its mixing; 3). During
all process of equilibration all of subsystems are considered as
equilibrium.

The first condition is necessary for introduction internal energy
into the description of dynamics of systems, as the parameter
which is necessary to correct describing an energy exchange
between IES. The second condition allows avoiding a problem of
redefinition IES due to mixing particles. Last condition is known
from thermodynamics. It removes the problems connected to the
description of system under condition of infringement of IES
equilibrium.

Below the substantiation of this approach is submitted. The UVS
equation and expression for dissipative force determining change
of internal energy IES are obtained. It is shown how basing on UVS
and D'Alambert principle, the obtaining of the generalized
Liouville equations for IES is possible. Formulas for entropy of
nonequilibrium systems are submitted. Connection between UVS and
the basic equation of thermodynamics is analyzed. Here we also
show how and why the offered approach has allowed to substantiate
thermodynamics within the framework of laws of classical
mechanics.

The problem of the description of nonequilibrium systems is a key
problem for physics as a whole. Therefore alongside with
mathematical calculations the discussion of the reasons why this
problem could not be solved on the basis of canonical Hamilton
equations is offered.

\section{A substantiation of the approach idea}

The classical mechanics is based on the abstract concept of the
elementary particle and the law of conservation of energy. Use of
these concepts allows constructing classical mechanics on the
basis of postulate: "Work of forces of reaction is always equal to
zero on any virtual displacement which is not breaking set
kinematics connections" [9]. Taking into account a condition of
conservatism of the active forces leads to a principle of the
least action, equations of Lagrange, Hamilton and Liouville
[9-12]. These equations describe systems' dynamics in equilibrium
states and near by. But all attempts to describe the dynamics of
nonequilibrium systems with their help collide with a serious
complexities. The analysis of these attempts allows assuming that
some restrictions used at creation of a mathematical formalism of
classical mechanics are unacceptable for the description of
evolution of nonequilibrium systems. {\it{Our task is to determine
these restrictions and to discover a way of construction of the
theory allowing their elimination.}}

The evolutionary processes in nonequilibrium system are caused by
internal forces and streams of energy created by them. These
streams is impossible to describe  with the help of a Hamilton
formalism. One of the reasons is impossibility of application of a
formalism of Hamilton for parts of system [9]. There is also other
not less important reason. The Hamilton formalism is constructed
on the basis of differential D'Alambert principle under condition
of conservatism of collective forces [9]. This requirement
excludes irreversibility automatically. But performance of a
condition of conservatism of forces is strictly proved only for
the equilibrium systems and in the approach of the theory of
disturbances. In all other cases including a case of
nonequilibrium systems, the strict proof of conservatism of
collective forces is absent [9-12]. {\it{Therefore it is necessary
to search for such approach to the analysis of nonequilibrium
systems which allows to exclude the using of the requirement of
potentiality of collective forces.}}

It was already mentioned that when the model of system of disks is
presented by plurality IES, the irreversibility can be shown by
analytical way. It is connected with non-potential collective
forces between IES changing their internal energy [5, 6]. It means
that {\it{for the analysis of nonequilibrium systems instead of
model of system consisting of elementary particles it is necessary
to use the model of system in the form of plurality of IES.}} For
such model the solution of a problem of irreversibility is reduced
to the proof of presence of non-potential forces between IES.

Necessity of splitting of system on a set of equilibrium
subsystems at the description of its evolution follows from a
statistical physics and a physical kinetics. In nonequilibrium
systems for each physical point there corresponds the local
velocity of a stream determining energy of a motion in this point.
Besides in this point the internal energy is exist. This energy is
determined by energy of the chaotic motions of particles. The
nature of these two energies is various. The first is caused by
transport of mass in the external field. The second energy is
determined by a chaotic motion of particles. Therefore the
distribution function at local equilibrium is determined by energy
of a motion of IES center of masses (CM) and energy of a chaotic
motion of particles [13]. The relative motion of IES disappears at
an establishment of equilibrium. I.e. energy of motion of IES is
the parameter describing a rate of a nonequilibrium.

It is possible to understand the necessity of splitting of IES
energy on two types: internal energy and energy of CM motion from
the analysis of two body system. As it is well known the problem
of two bodies could not be solved directly in a laboratory
coordinate system. It is related to the nonlinearity arising due
to interaction between bodies. But the problem can be easily
solved by transition into CM in which variables are parted. Such
transition is equivalent to representation of energy in the two
forms: energies of a motion of system as a whole and its internal
energy. In the absence of exterior forces the motion energy of
system is constant and a task is reduced to definition of the
relative motion of particles. The problem can be solved by
differentiation on time of the energy presented in the form of
motion energy and an internal energy.

{\it{So, we will build the approach to analyzing of the dynamics
of nonequilibrium systems within the frame of classical mechanics
by the next way. A requirement of absolutely elasticity of the
elements we will exclude by replacement of model of system of
elementary particles on model of the system consisting from
equilibrium structured particles, i.e. IES. Then, having presented
of the IES energy as the sum of energies of its motion and an
internal energy, we shall find the UVS. Using UVS the equations of
Lagrange, Hamilton and Liouville will be obtained without use of
the requirement of conservatism of collective forces.}}

\section{The system motion equation}

Let us obtain the motion equation for particle in an external
field. Let us presume that $E=m{v^2}/2+U(r)=const$ is a particle
energy. Here $m$ is a mass; $T=m{v^2}/2$ is a kinetic energy;
$U(r)$ is a potential energy. Then from eq. $\dot{E}=0$, we will
obtain:
\begin{equation}
v(m\dot{v}+\partial{U}/\partial{r})=0\label{eqn1}
\end{equation}
The eq. (1) is a balance equation of the kinetic and potential
energies. It is carried out if the condition takes place:
\begin{equation}
m\dot{v}=-\partial{U}/\partial{r}\label{eqn2}
\end{equation}
It is Newton equation (NE). This equation is integrable because a
variables were separated. The separation of variables became
possible in due to splitting of energy into kinetic and potential
components, each of which depends on the different variables. We
see that the sum of active and inertial forces is equal to zero.
The particle moves lengthways of a gradient of potential. The work
on the closed contour is equal to zero. Therefore the dynamics of
a particle is reversible.

Let us consider a system consists of $N$ potentially interacting
elements; the mass of each element is equal to 1.  The force
between any two elements is a central and determined by the
distance between them. Energy of system consist of the sum of
kinetic energy of elements, $T_N=\sum\limits_{i=1}^{N}
m{v_i}^2/2$, their potential energy in a field of external forces,
- ${U_N}^{env}$, and the potential energy of their interaction
${U_N}(r_{ij})={\sum\limits_{i=1}^{N-1}}{\sum\limits_{j=i+1}^{N}}U_{ij}(r_{ij})
$, where $r_{ij}=r_i-r_j$ - is a distance between elements $i$ and
$j$. So, $E=E_N+U^{env}=T_N+U_N+U^{env}=const$. The time
derivative of the energy will be as follows:
\begin{equation}
{\sum\limits_{i=1}^{N}}v_i\tilde{F}_{i}=0 \label{eqn3}
\end{equation}

Where $\tilde{F}_i=m\dot{v}_i+
\sum\limits_{j\neq{i}}^{N}F_{ij}+F_i^{env}$ is effective force for
$i$ particle; $\dot{U}^{env}=\sum\limits_{i=1}^{N}v_iF_i^{env}$;
$F_{ij}=\partial{U_N}/\partial{r_{ij}}$;
$F_{i}^{env}(r_i)=\partial{U^{env}}/\partial{r_{i}}$.

The eq. (3) can be treated as orthogonality of the vector of
effective forces with respect to the vector of velocities of
elements of the system. If there are no restrictions imposed on the
$v_i$ directions, the requirement $\tilde{F}_i=0$ is satisfied [9].
Then from eq. (3) we obtain:

\begin{equation}
{m\dot{v}_i=-\sum\limits_{i=1}^{N}}v_i\tilde{F}_{i}-F_i^{env}
\label{eqn4}
\end{equation}

It is NE for the system's elements in non-homogeneous space. It
leads to conclusion that the motion of an element of system is
determined by the force which equal to the sum of vectors of
forces, acting from all other particles and external force [11,
12].

Let us consider the motion of a system as the whole in a field of
external forces. As well as for an elementary particle its kinetic
energy is determined by the motion of CM. But except of this
energy the system possess of internal energy. This energy has
other nature because it is connected with interactions between
particles but not with the external field of forces. The internal
energy as well as energy of CM motion should vary due to the work
of the external forces. Really, the external field will make work
on change of internal energy due to motion of the particles
relative to the CM. But velocity of CM motion does not depend on
motion of elements inside of system. In connection with it, the
energy of system should be presented as the sum of energy of
motion of elements relative to the CM and the energy of CM motion.
According with such dividing the system's energy into two types,
the external forces also will be represented by the sum of two
forces: the force changing the velocity of CM and the force
changing the internal energy.

Let us take into account the equality: $T_N=\sum\limits_{i=1}^{N}
m{v_i}^2/2=M_NV_N^2+(m/N)\sum\limits_{i=1}^{N-1}\sum\limits_{j=i+1}^{N}v_{ij}^2$
(a), where $V_N=\dot{R}_N=1/N\sum\limits_{i=1}^{N}\dot{r}_i$ -are
velocities of the CM; $R_N$ - are coordinates of the CM;
$v_{ij}=\dot{r}_{ij}$.

Let's designate: $E_N=T_N^{tr}+E_N^{ins}$  where
$E_N^{ins}=T_N^{ins}+U_N$ is entrance energy, $T_N^{tr}$ is a CM
kinetic energy. The velocity of elements is $v_i=\tilde{v}_i+V_N$
where $\tilde{v}_i$ is a velocity of particle relative to the CM.
Then: $T_N=M_NV_N^2+mV_N\sum\limits_{i=1}^{N}\tilde{v}_i
+\sum\limits_{i=1}^{N}m\tilde{v}_{i}^2/2$. Because
$\sum\limits_{i=1}^{N}\tilde{v}_i=0$ then from (a) we have:
$\sum\limits_{i=1}^{N}m\tilde{v}_{i}^2/2=
1/(2N)\sum\limits_{i=1}^{N-1}\sum\limits_{j=i+1}^{N}v_{ij}^2$.
Thus the total kinetic energy of relative motion of particles is
equal to the sum kinetic energies of their motions relative to CM.
Because $r_{ij}=\tilde{r}_{ij}=\tilde{r}_i-\tilde{r}_j$, where
$\tilde{r}_i, \tilde{r}_j$ - are coordinates of the elements with
respect to the system's CM then
$U_N(r_{ij})=U_N(\tilde{r}_{ij})=U_N(\tilde{r}_i)$ and
$\sum\limits_{i=1}^{N-1}\sum\limits_{j=i+1}^{N}v_{ij}F_{ij}(r_{ij})
=\sum\limits_{i=1}^{N}\tilde{v}_iF_i(\tilde{r}_i)$, where
$F_i=\partial{U_N}/\partial{\tilde{r}_i}
=\sum\limits_{j\neq{i},j=1}^{N}\partial{U_N}/\partial{r_{ij}}$.
Thus we have [14]:
\begin{eqnarray}
V_NM_N\dot{V}_N+{\dot E}_N^{ins}=-V_NF^{env}-\Phi^{env}\label{eqn5}
\end{eqnarray}
Here $F^{env}=\sum\limits_{i=1}^{N}F_i^{env}(R,\tilde{r}_i)$, ${\dot
E}_N^{ins}=\sum\limits_{i=1}^{N}\tilde{v}_i(m\dot{\tilde{v}}_i+F(\tilde{r})_i)$,
$\Phi^{env}=\sum\limits_{i=1}^{N}\tilde{v}_iF_i^{env}(R,\tilde{r}_i)$.

The eq. (5) represents balance of energy of system in a field of
external forces. The first term in the left hand side determines
change of kinetic energy of system. The second term determines the
change of internal energy of system. The first term in the right
hand side determines the work of forces changing energy of motion
of system as a whole. The second term determines the work of
forces changing internal energy of system. This work is connected
with the motion particles of system relative to CM in the external
field. Thus the work of external forces changes $T_N^{tr}$ and
$E_N^{ins}$.

When the external forces are absent the eq. (5) will split on two
independent equations: the first is the equation of CM motion and
the second is the equation of motion of particles relative to the
CM.

Let us note that the eq. (5) can be obtained directly basing on
the NE for elements. For this purpose we shall multiply the eq.
(4) on the corresponding velocity. After summation the obtained
equations for all particles we shall have the eq. (5) (if we have
summarized the eq. (4) without multiplying it on velocity in this
case the internal forces in the second term of the eq. (5) will be
lost [15]). It is confirms validity of the equation (5).

Let us compare dynamics of a particle and dynamics of system. As it
follows from the eq. (1, 4) the force acting on the particle is
potential. The particles dynamics is determined by kinetic and
potential energies. Obviously {\it{it is impossible to find
particles' trajectory if not to divide the energy into the potential
and kinetic parts.}}

In agreement with the eq.(5) the work of external forces determining
motion of system goes both on changes of its CM velocity and on
change of internal energy. This force can be divided on two forces.
The first force is potential. The momentum of IES is change by this
force. The second force changes the internal energy. It is
non-potential force. Thus {\it{it is impossible to describe dynamics
of system if not to divide the energy of system on three types:
kinetic energy of motion of CM, internal energy and system's
potential energy in external field.}}

The equation of motion of system in an external field can be
obtained from the eq. (5). It may be expressed as:
\begin{eqnarray}
M_N\dot{V}_N= -F^{env}-[{\dot
E}_N^{ins}+\Phi^{env}]V_N/V_N^2\label{eqn6}
\end{eqnarray}
Let us call eq. (6) as generalized NE (GNE) for the structured
particle. The GNE is reduced to the NE if one neglects the
relative motion of elements, i.e. when the internal energy does
not change. In this case the dynamics of system is similar to
reversible dynamics of an elementary particle.

Thus, {\it{it is necessary to present the energy of a system as a
sum of energy of motion and internal energy for determination a
system dynamics in an external field. The external forces also
need to be splitting into the forces changing these types of
energies accordingly}}. Otherwise variables are not parted.

\section{The equation of interaction of two subsystems}

As it follows from the kinetic equations the dynamics properties
of the nonequilibrium systems' are determined by the energy of
motion and internal energy in each physical points of the system
[7, 13]. These parameters can be entered into the description of
dynamics of system if the system is presented by a set of IES or,
in other words, by a set of the structural particles. Using model
of system as a set of IES help us "select" nonequilibrium effects
into interactions of IES. These interactions can be determined by
the eq. (5) if external forces will be replaced by the forces
between IES. Let us show how it can be done.

Let suppose the system consists of two IES - $L$ and $K$. We take
all elements to be identical and have the same weight 1, and $L$
to be a number of elements in $L$ - IES, $K$ -is a number of
elements in $K$ -IES, i.e. $L+K=N$,
$V_L=1/L\sum\limits_{i=1}^{L}v_i$ and
$V_K=1/K\sum\limits_{i=1}^{K}v_i$ - are IES's velocities with
respect to the CM of system. The velocity of the system's CM we
take equal to zero, i.e. $LV_L+KV_K=0$.

We can represent the energy of the system as
$E_N=E_L+E_K+U^{int}=const$, where $E_L$ and $E_K$ are the IES
energy, and $U^{int}$ - is the energy of their interaction.
According to the eq. (6), the energy of each IES can be
represented as $E_L=T_L^{tr}+E_L^{ins}$, $E_K=T_K^{tr}+E_K^{ins}$,
where $T_L^{tr}={M_L}V_L^2/2$, $T_K^{tr}={M_K}V_K^2/2$, $M_L=mL,
M_K=mK$. $E^{ins}$- is the internal energy of a IES. The $E^{ins}$
consists of the kinetic energy of motion of the elements with
respect to the CM of IES - $T^{ins}$ and their potential energy -
$U^{ins}$, i.e. $E^{ins}=T^{ins}+U^{ins}$, where
$U_L^{ins}=\sum\limits_{i_L=1}^{L-1}\sum\limits_{j_L=i_L+1}^{L}U_{{i_L}{j_L}}(r_{i_Lj_L})$,
$U_K^{ins}=\sum\limits_{i_K=1}^{K-1}\sum\limits_{j_K=i_K+1}^{K}U_{{i_K}{j_K}}(r_{i_Kj_K})$.
The energy $U^{int}$ is determined as
$U^{int}=\sum\limits_{j_K=1}^{K}\sum\limits_{j_L=1}^{L}U_{j_Lj_K}(r_{j_Lj_K})$.
Indexes $j_k,j_L,i_K,i_L$ determine belonging of the elements to
corresponding IES. In equilibrium we have: $T^{tr}=0$. Hence, if
the system aspirates to equilibrium, then $T^{tr}$ energy for each
IES will be transformed into the internal energy of IES.

We can obtain the equations of dynamics of $L$ and $K$ of IES in
the following way. Let us differentiate energy of system on time.
In order to find the equation for $L$ - IES, at the left hand side
of obtained equality we have kept only those terms which determine
the change of the kinetic and potential energies of interaction of
elements of $L$ - IES. We replaced all other terms in the right
hand side and combined the groups of terms in such a way when each
group contains the terms with identical velocities. In accordance
with NE (see eq. (5), the groups which contain terms with
velocities of the elements from $K$- IES are equal to zero. As a
result the right hand side of the equation will contain only the
terms which determine the interaction of the elements  $L$-IES
with the elements $K$-IES. The equation for $K$-IES can be
obtained in the same way. Then we execute replacement of
variables: $ v_i=\tilde{v}_i+V$ and take into account equality
(a). As a result we will have [8, 14]:
\begin{eqnarray}
V_LM_L\dot{V}_L+{\dot{E}_L}^{ins}=-{\Phi}_L-V_L{\Psi}
\end{eqnarray}
\begin{eqnarray}
V_KM_K\dot{V}_K+{\dot{E}_K}^{ins}={\Phi}_K+V_K{\Psi}
\end{eqnarray}

Here $\Psi=\sum\limits_{{i_L}=1}^LF^K_{i_L}$;
${\Phi}_L=\sum\limits_{{i_L}=1}^L\tilde{v}_{i_L}F^K_{i_L}$;
${\Phi}_K=\sum\limits_{{i_K}=1}^K\tilde{v}_{i_K}F^L_{i_K}$;
$F^K_{i_L}=\sum\limits_{{j_K}=1}^KF_{i_Lj_K}$;
$F^L_{j_K}=\sum\limits_{{i_L}=1}^LF_{i_Lj_K}$;
${\dot{E}_L}^{ins}={\sum\limits_{i_L=1}^{L-1}}\sum\limits_{j_L=i_L+1}^{L}v_{i_Lj_L}
[\frac{{m\dot{v}}_{i_Lj_L}}{L}+\nonumber\\+F_{i_Lj_L}]$;
${\dot{E}_K}^{ins}={\sum\limits_{i_K=1}^{K-1}}\sum\limits_{j_K=i_K+1}^{K}v_{i_Kj_K}
[\frac{{m\dot{v}}_{i_Kj_K}}{K}+\nonumber\\+F_{i_Kj_K}]$.

The eqs. (7, 8) are the UVS. They describe energy exchange between
IES. Independent variables of UVS are macroparameters, coordinates
and velocities of IES motion, and also microparameters,
coordinates and velocities of elements. UVS binds together two
types of the description: at a macrolevel and at a microlevel. The
description at a macrolevel determines of IES dynamics as a whole
and at a microlevel determines dynamics of IES elements .

The potential force, $\Psi$, determines motion of IES as a whole.
This force is the sum of the potential forces acting on elements
of one IES at the side of another IES.

The forces which determined by the terms ${\Phi}_L$ and ${\Phi}_K$,
transformed of the motion energy of IES to internal energy as a
result of chaotic motion of elements of one IES in a field of the
forces of another IES.They are non-potential forces which dependent
on velocities and cannot be expressed as a gradient from any scalar
function. These forces are equivalents to dissipative forces.

The motion eqs. for IES can be obtain from eqs. (7,8):
\begin{equation}
M_L\dot{V}_L=-\Psi-{\alpha}_LV_L \label{eqn7}
\end{equation}
\begin{equation}
M_K\dot{V}_K=\Psi+{\alpha}_KV_K\label{eqn8}
\end{equation}
where ${\alpha}_{L}=(\dot{E}^{ins}_{L}+{\Phi}_{L})/V^2_{L}$,
${\alpha}_{K}=({\Phi}_{K}-\dot{E}^{ins}_{K})/V^2_{K}$,

The eqs. (9, 10) are GNE for interacting of systems. The second
terms in the right hand side of the equations determine the forces
changing internal energy of IES. These forces are equivalent to
friction force. Its work consists of works of chaotic motion of
elements one IES in a field of forces of another IES. The efficiency
of transformation of energy of relative motion of IES into internal
energy are determined by the factors "$\alpha_L$", "$\alpha_K$". If
the relative velocities of IES elements are equal to zero the force
of friction is also equal to zero.

For the description of dynamics of nonequilibrium systems we
instead of traditional model of system have taken model of system
as a set of interacting IES. Here a role of particles is carried
out by IES. Their motion is determined by the eqs. (9, 10).

The state of system from a set of IES can be determined by the
point in the phase space which consists of $6R-1$ coordinates and
momentum of IES, where $R$ is a number of IES. Let us call this
space as $S$-space to distinguish it from usual phase space for
elementary particles. Unlike usual phase space the $S$-space is
compressible though total energy of all elements is a constant.
The rate of compression of $S$-space is determined by the rate of
transformation of motion energy of the IES into their internal
energy. Thus the volume of compression of $S$-space is determined
by energy of the IES motion.

Let us compare GNE and NE. The NE follows from representation of
energy of an element as the sum of potential and kinetic energies.
The transformation of one energy type to another is determined
only by the potential force under condition of conservation of
full energy.

The GNE determines motion of IES not only at transformation of
kinetic and potential energies of IES, but also at transformation
of IES internal energy. These transformations are determined by
two forces. The potential part of force changes velocity of IES's
CM. The non-potential part of force changes its internal energy.
Thus, {\it{for the structured particle external force will consist
of two parts: potential and non-potential. Each of them has
collective character.}}

The GNE passes into NE when the relative velocities of elements
IES can be neglected. It is also have a place when distances
between IES are big enough [16].

\section{The generals of Lagrange, Hamilton and Liouville equations for IES}
The dynamics of the nonequilibrium system submitted by a set of
IES will be determined by generalized Lagrange, Hamilton and
Liouville equations for IES. Before explaining a way of obtain of
these equations, we shall remind under what conditions the
canonical principle of Hamilton for system of elementary particles
was deduced [9].

According to the differential principle of D'Alambert {\it{"the work
of effective forces including inertial and active forces, at virtual
motions elements of system is equal to zero for all reversible
virtual motions compatible with restrictions on dynamics"}} [9].
From D'Alambert principle with the help of NE the integrated
principle of Hamilton is deduced. For this purpose the integral on
time of the virtual work  made in system by effective forces is
equated to zero. Integration on time is carried out provided that
external forces possess power function. It means that the canonical
principle of Hamilton is fair only for cases when $\sum F_i\delta
R_i=-\delta U$ where $i$ is a number of particles, and $F_i$ - is a
force acting on this particle [9]. But for IES we have no right to
demand a carrying-out of a condition of conservatism of its forces.
From the mathematical point of view it is because the $\sum
F_i\delta R_i$ cannot be equal a full differential for IES. From the
physical point of view it is due to the non-potential force which
responds to the changes of the internal energy of IES. Therefore
obtaining of the Hamilton equations should be carried out basing on
GNE.

The principle D'Alambert for IES sounds so: {\it{the sum of works
of all forces of interaction IES at their virtual motions
compatible with restrictions on dynamics and in view of change of
internal energy of IES is equal to zero.}}

In according to the eqs. (9, 10) and Liouville formula from [6,
9], the generalized Liouville equations in $S$-space can be
written as:
\begin{equation}
df/dt=-f{\partial}{F}/{\partial}V \label{eqn11}
\end{equation}

Here $f$- is a distribution function for IES, $F$ -is a
non-potential part of  force acting on IES.

The right hand side of the eq. (11) is not equal to zero as forces
of IES interaction which transformed of the energy of its relative
motion in its internal energy, depend on velocities of elements.
The $S$-space which determined by the coordinates and velocities
of IES will be compressed. Thus system of IES will converge to
equilibrium.

The impossibility of return of internal energy of IES in its
energy of motion is caused by impossibility of change of IES
motion due to motion of its elements. At an establishment of
equilibrium the motion energy of IES will disappear. In this case
the description in $S$-space coincides with the traditional
description in usual phase space for elements of system.

\section{UVS and thermodynamics}

It is possible to come to the idea about necessity of replacement
of model of elementary particles by the model of interacting IES
from the analysis of the basic equation of thermodynamics.
According to this equation the work of external forces acting on
the system are splitting on two parts. The first part is connected
to reversible work. It can be associated with change of energy of
motion of system as a whole. Second part of energy will go on
heating. It is connected to internal degrees of freedom of system.
To this part of energy there corresponds internal energy of IES.
Here we would like to show, how basing on the UVS it is possible
to come to the thermodynamics from the classical mechanics.

Let us take the motionless nonequilibrium system consisting of
"$R$" of IES. Each of IES consists of enough plenty of elements.
Let $dE$ is a work which has been done at system. In
thermodynamics that term is an internal energy of a system (do not
confuse $E$ with the $E^{ins}$ - internal energy of IES). The $dE$
is determined by the basic equation of thermodynamics as:
${dE=dQ-PdY}$ [8]. Here, according to common terminology, $E$ is
energy of a system; $Q$ is thermal energy; $P$ is pressure; $Y$ is
volume.

As well as the basic equation of thermodynamics, UVS also is
differential of two types of energy. According to the UVS the
volume $dE$ is redistributed inside of the system in such a way
that that one its part goes on change of energy of relative motion
of IES, and another part changes its internal energy. The first
term in the left hand side of UVS is a change of kinetic energy of
motion of a IES as a whole, $dT^{tr}$. This term corresponds to
the term ${PdY}$. Really ,
${dT^{tr}=VdV=V\dot{V}dt=\dot{V}dr=PdY}$ [7].

If the potential energy is a homogeneous function of a second order
of the radius-vectors, then in agreement with the virial theorem
[10], we have:
${\bar{E}^{ins}=2\bar{\tilde{T}}^{ins}=2\bar{\tilde{U}}^{ins}}$. The
line denotes the time average.

Let us consider the system near to equilibrium state. The average
energy of each element is
${\bar{E}^{ins}={E}^{ins}/N=\kappa{T}_0^{ins}}$ where $N$ is a
number of elements. As the increasing of the internal energy is
determined by the volume ${dQ}$, then we will have:
${dQ\approx{T}_0^{ins}[d{E}^{ins}/{T}_0^{ins}]
\sim{T}_0^{ins}[{dv_0}/{v_0}]}$, where ${v_0}$ is the average
velocity of an element, and ${dv_0}$ is its change. For the system
in the closed volume we have:
${dv_0/v_0\sim{{d\Gamma}/{\Gamma}}}$, where ${\Gamma}$ is the
phase volume of a system, ${d\Gamma}$ will increase due to
increasing of the system's energy on the value, ${dQ}$. By keeping
the terms of the first order we get:
${dQ\approx{T}_0^{ins}d\Gamma/\Gamma={{T}_0^{ins}}d\ln{\Gamma}}$.
By definition ${d\ln{\Gamma}=dS^{ins}}$, where ${S^{ins}}$ is an
entropy [8]. So, near to equilibrium state we have
$d\bar{E}^{ins}={dQ\approx{T}_0^{ins}dS^{ins}}$.

The entropy increasing, $\Delta{S}$, for nonequilibrium system is
completely determined by the energy $T^{tr}$ passing into $E^{ins}$.
Therefore $\Delta{S}$ can be determined by the formula [7, 14]:

\begin{equation}
{{\Delta{S}}={\sum\limits_{L=1}^R{\{{N_l}
\sum\limits_{k=1}^{N_l}\int{\sum\limits_s{{\frac{{F^{L}_{ks}}v_k}{E^{L}}}}{dt}}\}}}}\label{eqn12}
\end{equation}

Here ${E^{L}}$ is the kinetic energy of $L$-IES; $N_L$ is the
number of elements in $L$-IES; $L=1,2,3...R$; ${R}$ is the number
of IES; ${s}$ is a number of the external element which
interaction with element ${k}$ belonging to the $L$-IES;
${F_{ks}^{L}}$ is a force, acted on $k$-element; $v_k$ -is a
velocity of the $k$- element.

In agreement with eq. (12), the entropy is determined by the energy
of the relative motion of IES, transformed in an internal energy as
a result of work of non-potential part of forces between IES.

To obtain equation for the entropy production we take into account
that: $\Delta{S}=\Delta{Q}/T$. Thus
$dS/dt=[dE^{ins}/dt]/(kE^{ins})$, where $k$ is a coefficient. It
is possible to express this formula through the work of forces of
interaction IES. Let $E_0$ is a full system's energy,
$E^{tr}_{L0}$ is a beginning energy of relative motion of $L$-IES.
In accordance with eqs. (10,11) the rate of increasing of the
internal system's energy is equal to: $\zeta=\sum^R_{L=1}\Phi_L$.
The internal energy of a system is equal to:
$E^{ins}=E_0-\sum^R_{L=1}E^{tr}_L$, where $\sum^R_{L=1}E^{tr}_L$
is a sum of IES energy of relative motion. But
$\sum^R_{L=1}E^{tr}_L=\zeta_0-\int^t_0\zeta(t){dt}$, where
$\zeta_0=\sum^R_{L=1}E^{tr}_{L0}$. Then entropy production for the
system, $\varrho_{prod}=dS/dt$, can be write:

$\varrho_{prod}=D/(1-D_0+\int^t_0D(t)dt)$, where $D=\zeta/E_0$,
$D_0=\zeta_0/E$.

The energy of $T^{tr}$ characterizes the rate of non-equilibrium
system while $E^{ins}$ characterizes its degree of equilibrium.

For maintenance of nonequilibrium system in a stationary state, loss
of energy $E^{tr}$ must be compensated by the inflow of external
energy. It is possible to create with the help of contact of system
with heater and due to outflow of radiation. Then the stationary
state of nonequilibrium system is characterized by the formula:
$\varrho_{prod}=|\varrho_{-}|-|\varrho_{+}|$, where $ \varrho_{-}$
is entropy outflow, $\varrho_{+}$ -is inflow of entropy,
$\varrho_{prod}$ is entropy production, determined by the formula
(12).

The process of transformation $E^{ins}$ to the energy of IES
motion is forbidden by the law of conservation of IES momentum
[10]. Really, momentum of IES cannot be changed by the forces
which change the internal energy. Therefore {\it{the work along
the closed contour for IES not equal to zero. As a result the
concept of entropy as the rate of increasing of IES internal
energy due to energy of its relative motion will appear within the
framework of classical mechanics. It is impossible to come to
entropy concept in the frame of classical mechanics if base on the
model of system as a set of elementary particles because for such
model the determining of $E^{ins}$ is impossible.}}

\section{Conclusion}
To connect thermodynamics with classical mechanics it is necessary
to refuse from the idealization of an elementary particle and a
condition of conservatism of collective forces. It can be made by
replacement of system of elementary particles on a system of IES.

Collective forces between IES are consisting of two parts. The
potential part of force changes energy of IES motion. The
non-potential part of force changes internal energy. The change of
internal energy is caused by chaotic motion of particles of one IES
in a field of forces of others IES. Therefore at equality to zero of
velocities of particles in relative to CM of IES, the changes of its
internal energy do not occur.

The Lagrange, Hamilton and Liouville equations on the basis of UVS
are deduced. These equations describe dynamics of the
nonequilibrium system in the compressible $S$-space where points
are corresponds to the momentum and coordinates of the IES. We
call this space as $S$-space in accordance with the entropy
designation. Compressing of $S$-space is explained by the
transition of energy of IES motion into their internal energy.

The following mechanism of irreversibility can be proposed: the
energy of relative motion of IES is transformed into their
internal energy as a result of the work of the non-potential part
of a force between of IES. The system equilibrates when relative
motion of IES disappears.

The explanation of the First law of thermodynamics is based on the
fact that the work of subsystems' interaction forces changes both
the energy of their motion and their internal energy.

The explanation of the Second law of thermodynamics is based on
the condition of irreversible transformation of the subsystems'
relative motion energy into their internal energy.

Thus within the limits of the accepted restrictions the offered
approach has allowed to expand a formalism of classical mechanics
on nonequilibrium systems and to connect it with thermodynamics.

Difficulties on a way of the solution of a problem of
irreversibility are resolved at refusal of the assumption of
absolute elasticity of particles and potentiality of collective
forces between IES.

\smallskip


\medskip
\begin{thebibliography}{9}

\bibitem{Ref1}
 Cohen E.G, Boltzmann and statistical mechanics, Dynamics: Models
and Kinetic Methods for Nonequilibrium Many Body systems.1998 NATO
Sci. Series E: Applied Sci., 371, p. 223

\bibitem{Ref2}
 Prigogine I, From the being to becoming. Moscow, 1980

\bibitem{Ref3}
Zaslavsky G.M, Chaotic dynamic and the origin of Statistical
laws,1999 Physics Today. August. Part 1, p.39

\bibitem{Ref4}
 Ulenbek G., Fundamental problems of the stat. mechanics, 1971 UFN.
V.103, N27, p.275

\bibitem{Ref5}
 Somsikov V.M, Nonrecurrence problem in evolution of a hard-disk
system, 2001 Intern. Jour. Bifurcation And Chaos 11, 11. 2863

\bibitem{Ref6}
 Somsikov V.M, Some approach to the Analysis of the Open
Nonequilibrium systems, 2002 AIP. 20, p.149

\bibitem{Ref7}
 Landau L.D,  Lifshits Ye. M., Statistical Physics. 1976 Nauka, Moscow,
583 p.

\bibitem{Ref8}
 Somsikov V.M., Thermodynamics and classical mechanics. Journal of
physics: 2005 Conference series, 23, p.7

\bibitem{Ref9}
 Lanczos C., The variation principles of mechanics. Univer. of
Toronto press. 1962, 408 p.

\bibitem{Ref10}
 Landau L.D,  Lifshits Ye.M., Mechanics. Nauka, Moscow, 1958, 115
p.

\bibitem{Ref11}
 Arnold V.,  Mathematical methods in classical mechanics. Mir,
Moscow, 1976, 431p.

\bibitem{Ref12}
Goldstein G, Classical mechanics. Ì., 1975, 416 p.

\bibitem{Ref13}
Rumer Yu.B, Ryvkin M.Sh, Thermodynamics. Stat. Physics and
Kinematics. M., Nauka, 1977, 532 p.

\bibitem{Ref14}
Somsikov V.M, Expansion of a formalism of classical mechanics for
nonequilibrium systems. arX:physics/0703141v1 17Mar2007.

\bibitem{Ref15}
Longmair K, Plasma physics. Atomizdat. Ìoscow, 1966, 341 p.

\bibitem{Ref16}
 Poincare A., About science. Ì.-Sci. Rus. 1983, 560 p.

\end{thebibliography}
\end{document}